\documentclass[Times,twocolumn,tighten,deluxetable]{aastex63}

\usepackage{graphicx}
\usepackage{rotating}
\usepackage{lineno}
\usepackage{multirow}
\accepted{on Sept. 5, 2024}
\submitjournal{ApJ}

\shorttitle{Detection of QPO Lag during Outburst of Swift J1727.8-1613}
\shortauthors{Debnath et al.}

\begin{document}

\title{Detection of QPO Soft Lag during Outburst of Swift J1727.8-1613: Estimation of Instrinsic Parameters from Spectral Study}

\correspondingauthor{Dipak Debnath ; Sujoy Kumar Nath}
\email{dipakcsp@gmail.com ; sujoynath0007@gmail.com}

\author[0000-0003-1856-5504]{Dipak Debnath}
\affiliation{Institute of Astronomy Space and Earth Science, P 177, CIT Road, Scheme 7m, Kolkata 700054, India}
\affiliation{Institute of Astronomy, National Tsing Hua University, Hsinchu 300044, Taiwan}

\author[0000-0002-6640-0301]{Sujoy Kumar Nath}
\affiliation{Indian Center for Space Physics,  466 Barakhola, Netai Nagar, Kolkata 700099, India}

\author[0000-0001-6770-8351]{Debjit Chatterjee}
\affiliation{Institute of Astronomy, National Tsing Hua University, Hsinchu 300044, Taiwan}

\author[0000-0002-6252-3750]{Kaushik Chatterjee}
\affiliation{South-Western Institute For Astronomy Research, Yunnan University, University Town, Chenggong, Kunming 650500, China}
\affiliation{Institute of Astronomy Space and Earth Science, P 177, CIT Road, Scheme 7m, Kolkata 700054, India}

\author[0000-0002-5617-3117]{Hsiang-Kuang Chang}
\affiliation{Institute of Astronomy, National Tsing Hua University, Hsinchu 300044, Taiwan}
\affiliation{Department of Physics, National Tsing Hua University, Hsinchu 300044, Taiwan}

%==================================================================================================================================

\begin{abstract}
The recently discovered bright transient black hole candidate Swift J1727.8-1613 is studied in a broad energy range ($0.5-79$\,keV) using 
combined NICER and {\it NuSTAR} data on 29 August 2023. A promonient type-C Quasi-Periodic Oscillation (QPO) at $0.89 \pm 0.01$\,Hz 
with its harmonic was observed in NICER data of $0.5-10$\,keV. Interestingly, the harmonic becomes weaker in the lower energy bands 
($0.5-1$ \& $1-3$\,keV). We also report the first detection of a soft time-lag of $0.014 \pm 0.001$\,s at the QPO frequency between 
harder ($3-10$\,kev) and softer ($0.5-3$\,keV) band photons observed with the NICER/XTI instrument. This indicates that the inclination 
of the accretion disk in the binary system might be high. From the detailed spectral analysis with the {\tt relxill} reflection model, 
we found the disk inclination angle of source to be $\sim 85^\circ$. We discuss how the accretion flow configuration inferred from 
spectral analysis can help us to understand the origin of QPOs and soft lag in this source.

\end{abstract}


\section{Introduction}

The spectral and timing properties of a low mass Black hole X-ray binary (BHXRB) show a rich variability during an outburst.
Two main states are defined based on these variations: the hard state (HS) and the soft state (SS). In the HS, the energy spectrum is dominated by a high energy Comptonized component described by a hard power law. 
On the other hand, the energy spectrum in the SS is dominated by a low energy thermal component described by a multicolour disk blackbody.
Between the HS and SS, a hard-intermediate state (HIMS) and a soft-intermediate state (SIMS) can be defined with properties in between the HS and SS (Homan \& Belloni 2005, Nandi et al. 2012).

Along with the long term variation in the X-ray fluxes and spectral nature, the X-ray emission from BHXRBs also show rapid variability with a timescale from milliseconds to seconds. 
In most of the BHXRBs, the power-density spectra (PDS) obtained from the short timescale light-curves show the presence of peaked and narrow noise components called quasi-periodic oscillations (QPOs). 
Their properties (centroid frequency, Q-value, rms amplitude etc.) vary during different spectral states.
Low frequency QPOs (LFQPOs) that have centroid frequencies ranging from a few mHz to 30\,Hz can be classified into three groups: type A, B, and C (Casella et al. 2005).
In the HS and the HIMS, the PDS is characterized by a strong and narrow type-C QPO peak with a Q-value of $\sim 7-12$ and an amplitude of $\sim3-16~\%$ rms along with a strong broad-band noise component.
Often a sub harmonic and a second harmonic peak are also present in the PDS in these states.
In the SIMS, the broad-band noise becomes weaker and a type-B QPO with Q-value of $\sim 6$ and rms amplitude of $\sim2-4~\%$ or a broader type-A QPO with Q-value of $\sim 3$ and rms amplitude of $\sim3~\%$ can be detected in the PDS (Casella et al. 2005).

Although observed frequently, the origin of the LFQPOs is a matter of a long standing debate.
The properties of these LFQPOs, such as the centroid frequencies and rms amplitudes often exhibit a correlation with the proportional dominance of thermal and non-thermal fluxes during different outburst states (Muno et al. 1999; Vignarca et al. 2003; Titarchuk \& Fiorito 2004).
It was shown that the Comptonized photons are responsible for the production of LFQPOs (Chakrabarti \& Manickam 2000; Rao et al. 2000; Vadawale, Rao \& Chakrabarti 2001; Chakrabarti et al. 2005), hence they are related to the properties of the Comptonizing region.
The time delay due to Compton up-scattering and down-scattering in this Compton cloud was proposed initially to be the mechanism that produced hard and soft lags associated to these LFQPOs (Miyamoto et al. 1988; Reig et al. 2000).
Time/phase lag between X-rays of different energy bands can be measured using their cross-spectrum, and it is a measure of the difference in time of arrival between photons in these energy bands.
By general convention, if the hard X-rays lag behind soft X-rays, the lag is defined as positive or hard lag.
On the contrary, if soft X-rays lag behind hard X-ray photons, the lag is defined as negative or soft lag.
Over the years, several models have been proposed to explain the hard and soft lags associated with LFQPOs in BHXRBs (Lee \& Miller 1998; Poutanen \& Fabian 1999; Ingram et al. 2009; Shaposhnikov 2012; Misra \& Mandal 2013 etc.).
Although some sources show hard lags for all frequencies, it was found that for some sources like XTE J1550-564 and GRS 1915+105 the lag transitions from hard to soft at around the QPO frequency (Qu et al. 2010).
From a survey of over 15 BHXRBs, van den Eijnden et al. (2017) found that soft lags are observed only for sources having a high disk inclination with respect to the observer's line of sight.
It was shown that multiple physical mechanisms, e.g., Compton up and down scattering, reflection of hard X-rays from the disk, gravitational bending etc. contribute differently to produce different kinds of lag behavior for low and high inclination sources (Dutta \& Chakrabarti 2016, Chatterjee et al. 2017a).
It was also observed that the presence of soft lags is concurrent with higher radio activity in high inclination sources, which indicates outflows can also influence the lag behavior (Patra et al. 2019, Chatterjee et al. 2020).

The Galactic transient black hole candidate (BHC) Swift J1727.8-1613 was first discovered by {\it Swift}/BAT as a Gamma-Ray Burst object 
(GRB 230824A) on 24 August 2023 (Kennea \& Swift Team 2023). The follow-up observations in X-rays (Negoro et al. 2023) and optical (Castro-Tirado et al. 2023) 
wavebands suggested that the source is a black hole low mass X-ray binary. During the outburst, the maximum flux of this bright source was found to be $\sim 7.6$ Crab 
(Palmer \& Parsotan 2023), which generated interest in the scientific community 
to observe it in multi-wavelength bands. It was monitored in a multi-wavelength basis, such as X-rays 
(Negoro et al. 2023; O'connor et al. 2023; Sunyaev et al. 2023; Dovciak et al. 2023; Katoch et al. 2023), optical (Castro-Tirado et al. 2023; 
Wang \& Belim 2023; Sanchez \& Munoz-Darias 2023), NIR (Baglio et al. 2023) and radio (Miller-Jhones et al. 2023; Trushkin et al. 2023; 
Patra et al. 2023) wavebands. This source showed a prominent signature of low-frequency QPOs (Palmer \& Parsotan 2023; Draghis et al. 2023; Katoch et al. 2023; 
Debnath et al. 2023) during the rising phase of the outburst. Debnath et al. (2023) reported a preliminary detection of a soft time-lag and 
a reflection dominated spectral signature.
Polarized emission was discovered in sub-mm wavelengths with the \textit{IXPE} satellite having a time 
and energy-averaged polarization degree of $4.1\% \pm 0.2\%$ and an angle of $2.2^\circ \pm 1.3^\circ$ (Veledina et al. 2023).
This detection indicates that the Compton corona is elongated in the orthogonal direction of the jet emission (Veledina et al. 2023).
A strong continuous jet has been observed in this source from polarization and radio analysis which could indicate the presence of the largest continuous jet observed in any X-ray binary (Wood et al. 2024).

We study the spectral and temporal properties of Swift J1727.8-1613 on 29th Aug 2023 using NICER/XTI and {\it NuSTAR}/FPM data. 
We report on the observations, data reduction, and analysis procedures briefly in \S2. In \S3 we present the results of our 
analysis. In \S4, we discuss the obtained results and present final conclusions.

\vskip 1.0cm

\section{Observation and Data Analysis}

\subsection{Observations}

We study the combined NICER and {\it NuSTAR} archival data of Swift J1727.8-1613 taken on 2023 Aug 29 (MJD= 60185). NICER/XTI data 
% of observation ID 6203980105 
are used for both temporal and spectral analysis in the energy range $0.5-10$\,keV, while {\it NuSTAR}/FPMA data are used only for spectral analysis. The combined NICER plus 
{\it NuSTAR} data allow us to study the nature of the source in a broad energy band of $1-70$\,keV.

\subsection{Data Reduction}

\subsubsection{NICER}
NICER is an external payload attached to the International Space Station which has an X-ray timing instrument (XTI; Gendreau et al. 2012) 
working in the energy range 0.2-12\,keV with a time resolution of $\sim0.1$ ${\mu}$s and spectral resolution of $\sim85$ eV at 1\,keV. 
For the analysis, the Level 1 data files are processed with {\fontfamily{qcr}\selectfont nicerl2} script in the latest caldb environment (ver. xti20221001) to obtain fully calibrated Level 2 event files that are screened for non-X-ray events or bad data times.
Light-curves with different time-bins in different energy ranges are then obtained with the task {\fontfamily{qcr}\selectfont nicerl3-lc}\footnote{https://heasarc.gsfc.nasa.gov/docs/nicer/analysis\_threads/nicerl3-lc}. 
To obtain the spectra, we use the task {\fontfamily{qcr}\selectfont nicerl3-spect}\footnote{https://heasarc.gsfc.nasa.gov/docs/nicer/analysis\_threads/nicerl3-spect}.

\subsubsection{NuSTAR}
{\it NuSTAR} raw data from the web archive are reduced with the {\it NuSTAR} data analysis software ({\fontfamily{qcr}\selectfont NuSTARDAS}, version 1.4.1).
Cleaned event files are produced using the {\fontfamily{qcr}\selectfont nupipeline} task using the latest calibration files.
With the {\fontfamily{qcr}\selectfont XSELECT} task, a circular region of $60''$ centred at the source coordinates is chosen as the source region, 
and a circular region with the same radius away from the source location is chosen as the background region.
The {\fontfamily{qcr}\selectfont nuproduct} task is then used to extract the spectrum, ARF and RMF files. 
The extracted spectra are then rebinned to have at least 30 counts/bin with the {\fontfamily{qcr}\selectfont GRPPHA} task.

\subsection{Data Analysis}

We use the Stingray software package (Huppenkothen et al. 2019) for temporal analysis.
We construct power density spectra (PDS) from data segments of 81.92\,s of the 0.01\,s time binned light-curve. 
Then we average all the individual power spectra and subtract the Poisson noise according to the average power in the 30-50\,Hz frequency range.
The resulting PDS is then geometrically rebinned.
We repeat the above procedure for 0.5-10\,keV, 0.5-1\,keV, 1-3\,keV and 3-10\,keV energy bands to generate PDSs of different energy bands. 
We model the PDS with multiple Lorentzian models in {\fontfamily{qcr}\selectfont XSPEC} version 12.13.0c (Arnaud 1996) to account 
for the broadband noise, the QPO peak and its harmonic.
From the fits we obtain the centroid frequencies ($\nu_{0}$), width ($\Delta\nu$), Q-value ($Q=\nu_{0}/\Delta\nu$) and RMS ($\%$) amplitude. 
We calculate the cross-spectrum between 0.5-3\,keV and 3-10\,keV light-curves. 
From the argument of the complex cross-spectrum, we obtain frequency dependent phase-lag and time-lag spectrum.
Time lags at the individual noise peaks are then obtained by averaging the time lag values over the width ($\Delta\nu$) 
of the Lorentzian component around the centroid frequency ($\nu_{0}\pm\frac{\Delta\nu}{2}$; Reig et al. 2000).

% We utilize HEASARC's spectral analysis software package {\fontfamily{qcr}\selectfont XSPEC} version 12.11.1 (Arnaud 1996)
% for analyzing the spectra. 

\section{Results}

\subsection{Energy dependent PDS}
We use 0.01\,s time-binned NICER/XTI light curves to construct the PDS in the full $0.5-10$\,keV energy band.
A prominent and narrow QPO peak was observed around 1\,Hz. To study the energy dependence of the noise components,
we extracted light-curves of $0.5-1$, $1-3$ and $3-10$\,keV energy bands and constructed the respective PDSs.
These are shown in the upper panel of Figure 1. As can be seen from the figure,
though the main QPO peak is present in all the PDSs, a harmonic of the main peak at $\sim2$\,Hz appears in the
$3-10$\,keV PDS. Furthermore, though the $1-3$\,keV PDS shows signs of the harmonic, it is entirely
absent in the $0.5-1$\,keV PDS. To determine the properties of the QPO and its harmonic, we fitted the $3-10$\,keV band 
PDS with multiple broad and narrow Lorentzian models (Nowak 2000, Belloni et al. 2002), which is shown in the lower panel in Fig. 1. From the fit, we find that the main 
QPO peak has a centroid frequency of $0.89^{+0.01}_{-0.01}$\,Hz with a Q value of $5.14$ and fractional RMS variability 
of $8.07\%$, whereas the harmonic peak has a centroid frequency of $1.78^{+0.03}_{-0.01}$\,Hz with a Q value of $4.25$ 
and fractional RMS variability of $5.26\%$. These values suggest that the observed QPO is of type-C (Casella et al. 2005).

\begin{figure}[!h]
%\vskip 0.5cm
  \centering
    \includegraphics[angle=0,width=9.2cm,keepaspectratio=true]{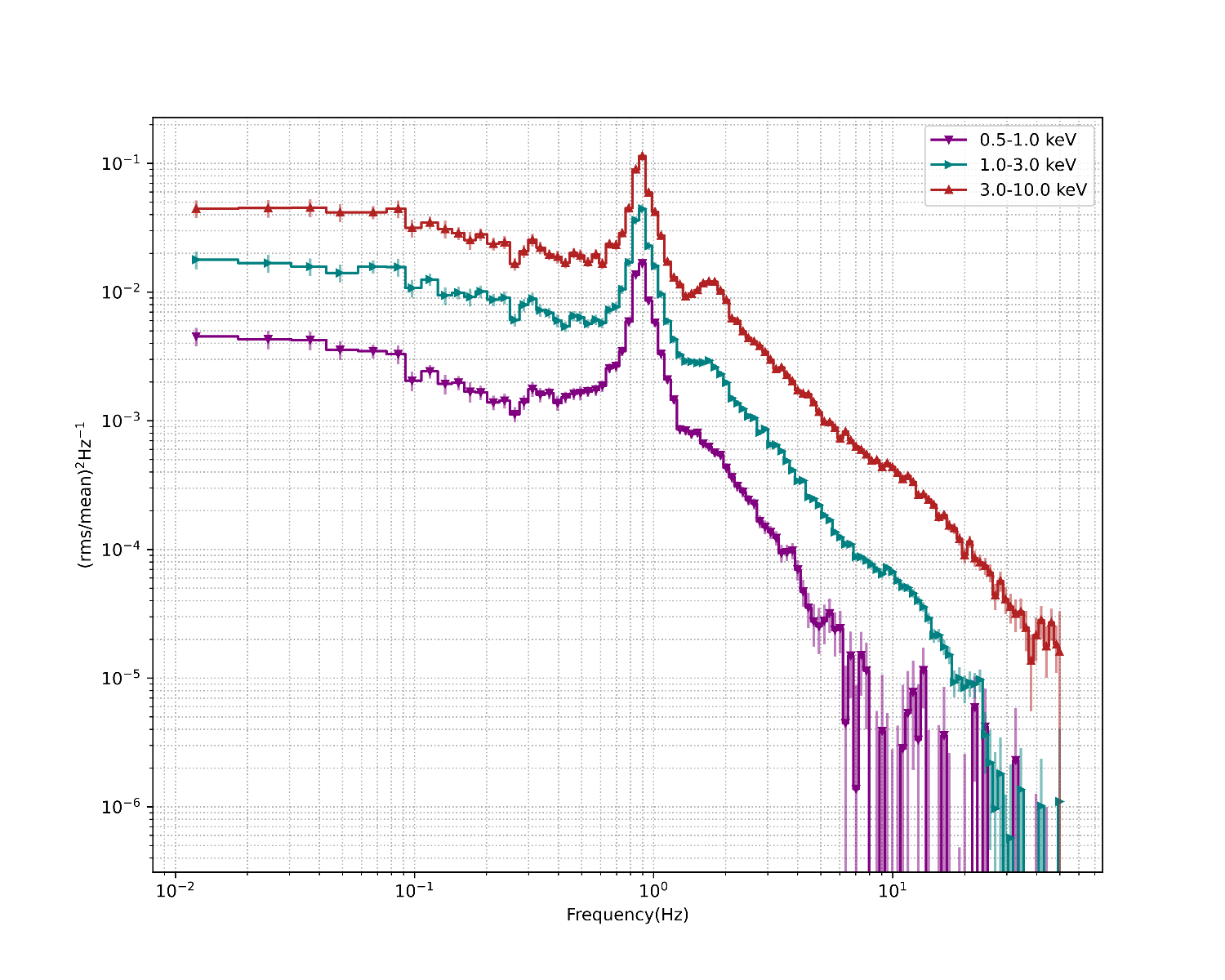}\hskip 0.02cm
    \includegraphics[angle=0,width=8cm,keepaspectratio=true]{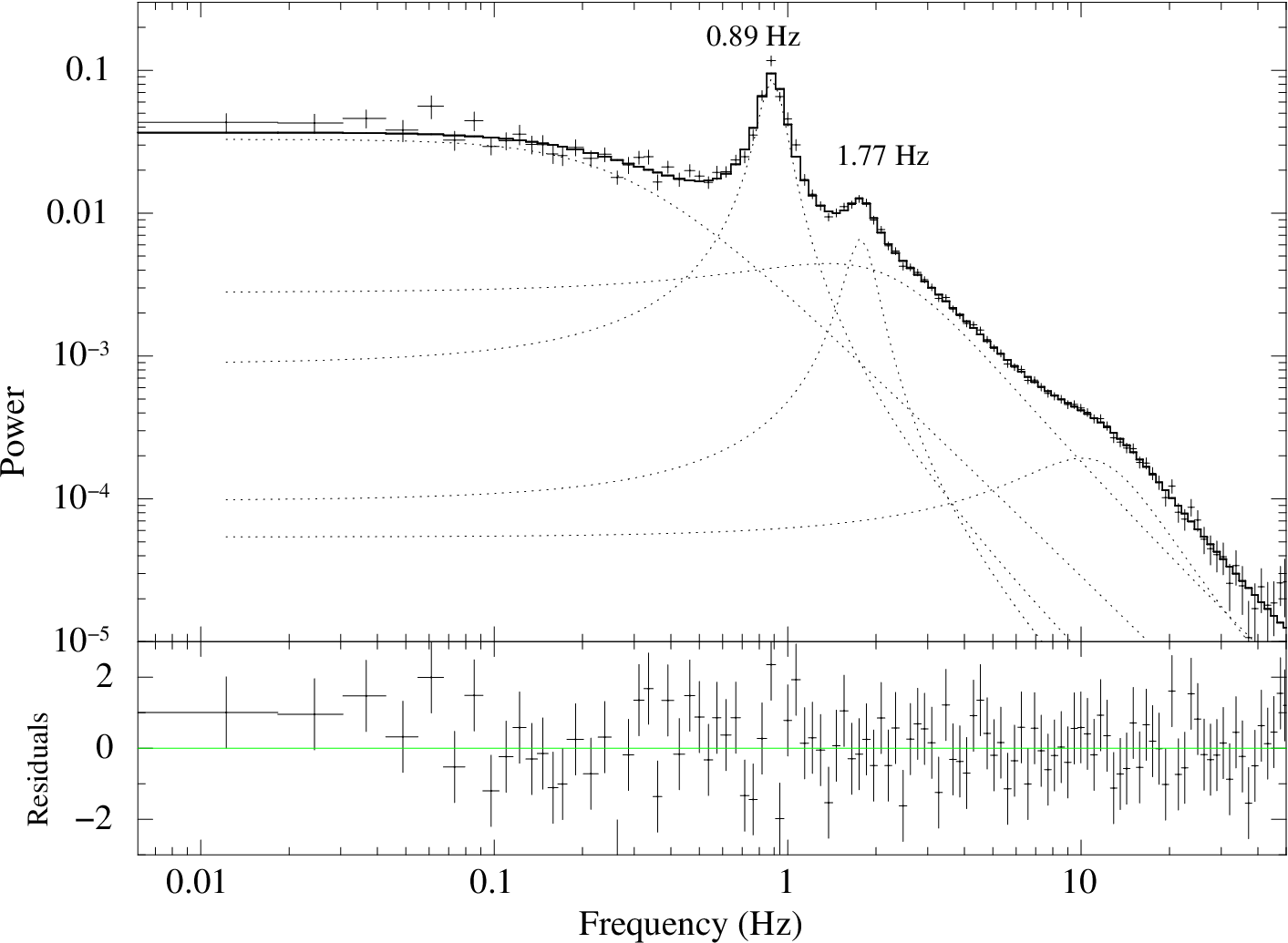}
	\caption{(a) The upper panel shows fast-Fourier transformed power-density spectrum (PDS) in four different energy bands ($0.5-1$, $1-3$, and
            $3-10$\,keV) of NICER. 
            (b) The lower panel shows the $3-10$\,keV PDS fitted with multiple Lorentzian models. A prominent type-C 
            QPO of $0.89$\,Hz and its harmonics is clearly detected.}
            
\end{figure}

\subsection{Time and Phase lag spectra}

\begin{figure*}%[!h]
\vskip -0.2cm
  \centering
    \includegraphics[angle=0,width=17cm,keepaspectratio=true]{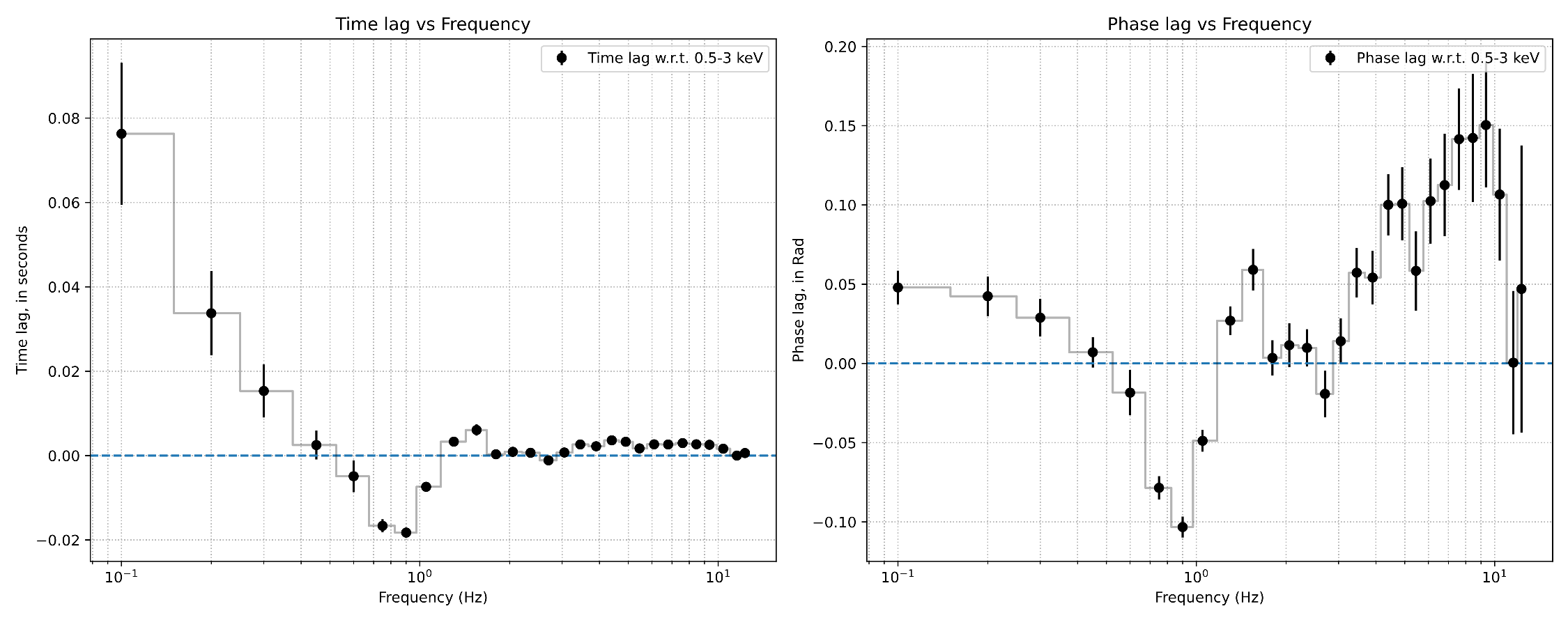}
	\caption{(a) Time lag and (b) Phase lag plots of Swift J1727.8-1613 on 2023 Aug 29 plots using  NICER light curves of soft ($0.5-3$\,keV) 
             and hard ($3-10$\,keV) band photons. Soft time lag of $0.014 \pm 0.001$\,s at the QPO frequency is found.
             }
\end{figure*}

We generate frequency dependent phase and time lag spectra between the $0.5-3$\,keV and $3-10$\,keV energy band light-curves 
following Vaughan \& Nowak (1997) and Nowak et al. (1999). These are shown in Figure 2. From the figure, we can see
positive lags at lower frequencies, implying the variabilities in the hard band ($3-10$\,keV) light-curve lag behind those 
of the soft band ($0.5-3$\,keV) at these frequencies. However, the lag spectrum switches sign at $\sim0.5$\,Hz 
and becomes negative showing maximum negative lag around the QPO fundamental frequency and then again becomes positive 
at around the frequency of the QPO harmonic. To calculate the time lags associated with the QPO and its harmonic, we 
averaged the lags over the width ($\Delta\nu$) of the corresponding Lorentzian component around its centroid frequency 
($\nu_{0}\pm\frac{\Delta\nu}{2}$) following Reig et al. (2000). The calculated time lag at the QPO frequency is $-0.014 \pm 0.001$\,s
implying a soft lag and at the QPO harmonic is $0.002 \pm 0.001$\,s implying a hard lag.

\subsection{Spectral Properties}

\begin{figure}
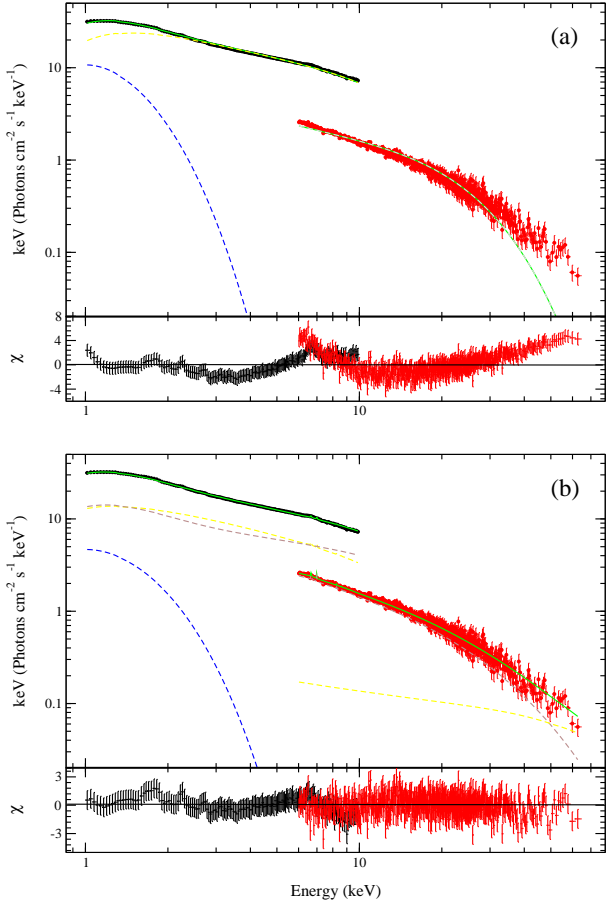
%[!h]
\vskip 0.5cm
  \centering
    \includegraphics[angle=0,width=8cm,keepaspectratio=true]{fig3a.eps}\hskip 0.2cm
    \includegraphics[angle=0,width=8cm,keepaspectratio=true]{fig3b.eps}
	\caption{Combined NICER (1-10\,keV) plus {\it NuSTAR} (6-70\,keV) spectrum fitted with $tbabs(diskbb+thcomp*diskbb)$ is shown in the top panel (a) 
            and in the bottom panel (b), combined spectrum is fitted with $tbabs(diskbb+{\tt xillver}+thComp*diskbb)$. Model fitted parameters are shown in Table 1. The black and red represent the NICER and {\it NuSTAR}/FPMA spectrum (or residuals), respectively. The blue dashed line indicates the $diskbb$ model component. The $thComp*diskbb$ model component is shown by the yellow dashed line, while the magenta color indicates the {\tt xillver} component in the bottom panel (b). The green dashed lines represent the total best-fit model.
      		}
\end{figure}

\begin{figure}
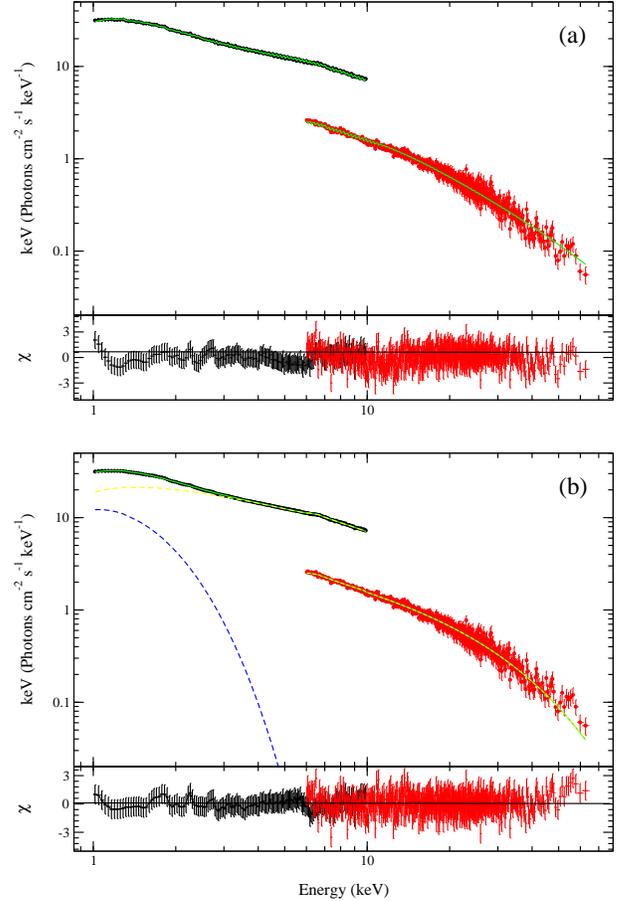
%[!h]
\vskip 0.2cm
  \centering
    \includegraphics[angle=0,width=8cm,keepaspectratio=true]{fig4a.eps}\hskip 0.2cm
    \includegraphics[angle=0,width=8cm,keepaspectratio=true]{fig4b.eps}
	\caption{Combined NICER (1-10\,keV) plus {\it NuSTAR} (6-70\,keV) spectrum fitted with (a) $tbabs({\tt reflect}*TCAF)$ (b) $tbabs(diskbb+{\tt relxill})$. In Table 1, model fitted parameters are shown. The black and red represent the NICER and {\it NuSTAR}/FPMA spectrum (or residuals), respectively. In Fig. (b), the blue dashed line indicates the $diskbb$ model component. The {\tt relxill} model component is shown by yellow dashed line. The green dashed lines represent the total best-fit model.
         }
\end{figure}

We fit the combined NICER and {\it NuSTAR}/FPMA data with a multi-color disk blackbody (Mitsuda et al. 1984; Makishima et al. 1986) and $thComp$ model in xspec. The $thComp$ model assumes a spherical source to comptonize the photons by hot electrons. It is an improved model with respect to the $nthComp$ model that describes the continuum shape of the the thermal Comptonisation spectra (Zdziarski 1996, 2020). It is used as a convolution model, and thus it Comptonizes both the soft and hard photons. The model parameters are similar to an exponential cut-off power-law model, although it gives a sharper cut-off than the exponential one. The high energy cut-off is described by an electron temperature ($kT_e$) in this case. The positive values of the photon index ($\Gamma_\tau$) give the low energy power-law photon index, while the negative value represents the absolute optical depth.
We also used $Tbabs$ for the absorption by interstellar medium, with `vern' cross-section (Verner et al. 1996) and `wilms' abundances (Wilms et al. 2000). Thus, our combined model is: const*Tbabs(diskBB+ThComp*diskbb). The `const' is used to combine the NICER and {\it NuSTAR} spectra with the same model normalization. The value of the `const' is fixed to one for NICER spectrum, while it is kept free to vary for {\it NuSTAR} data. We obtain an inner disk temperature of $kT_{\rm in}\sim$ 0.31$\pm$0.01\,keV and photon index of $\Gamma\sim$ 1.76$\pm$0.004. The electron temperature gives very low value ($kT_e\sim$ 5.63\,keV). The best-fit spectrum using this model gives a $\chi^2_{red}$ value of $\sim$2.86. In Fig. 3a, we show the fitted spectrum with the $\chi^2$ variation. Significant residuals can be seen around 5-8\,keV and after 40\,keV. There are also some small residuals at low energies, which have been seen in the other NICER observations (e.g. Draghis et al. 2023), and are considered to be instrumental effects. We included multiple Gaussian (model) components to fit the residuals, but the statistics did not improve significantly.
Hence we did not add any further Gaussian model components to fit those residuals since it may cause overfitting the whole spectra.
We include the {\tt xillver} model component (Garc\'ia \& Kallman 2010, Garc\'ia et al. 2013) in the previous composed model to consider the remaining residuals. The best-fit results are roughly similar to the previous combination with $kT_{\rm in}$ ($\sim0.39$\,keV) and $\Gamma$ ($\sim1.53$). Although, the $kT_e$ increases to 20.06$\pm$6.31\,keV. During the fitting we first tied the $\Gamma$ parameter of {\tt xillver} to the value of $\Gamma_\tau$ of thComp. After getting a best fit, we make it free to vary and again do the fit. The $\Gamma$ of {\tt xillver} gives a similar value to the $\Gamma_\tau$ of $thComp$. We fixed the iron abundance to the solar value. We get a high energy rollover at around 17.13$\pm$4.33\,keV. We also obtain the ionization parameter to be log$\xi\sim$ 3.62$\pm$0.09 and the inclination of the inner accretion disk as 79.99$^\circ$$\pm$11$^\circ$. The inclusion of the {\tt xillver} model component gives a satisfactory fit with $\chi^2_{red}\sim0.87$. The best-fit spectra are shown in Fig. 3b, and the fitted parameters are presented in Table. 1.
We conclude that the best description of the spectral features is a model composed by a multicolor disk blackbody component, a power-law component, and reflection.
%It is evident that the spectral features consist of a multicolor disk blackbody component, a power-law component, and reflection.

To understand the dynamic nature of the accretion flow, we modeled the combined spectra with the Two Component Advective Flow (TCAF; Chakrabarti \& Titarchuk 1995) model.
The TCAF model is a physical model based on the transonic accretion flow solutions of Chakrabarti (1990) and uses four main flow parameters to describe the dynamics of the accretion flow: the Keplerian disk and sub-Keplerian halo accretion rates ($\dot{m}_d$ \& $\dot{m}_h$ in $\dot{M}_{\rm Edd}$ unit), the location of the shock ($X_s$ in Schwarzschild radius $r_s=2 GM_{\rm BH}/c^2$) and the shock strength ($R = \rho_+/\rho_-$, ratio of the post-shock to the pre-shock matter density).
Additionally, the mass of the BH ($M_{\rm BH}$ in $M_\odot$ unit) and the model normalization ($N$) are required to fit the spectrum.
To account for the strong reflection features, we convolve this model with the {\tt reflect} model that describes reflection from neutral material (Magdziarz \& Zdziarski 1995). Thus the combined model becomes: const*Tbabs({\tt reflect}*TCAF). Using this model, we get a satisfactory fit with $\chi^2_{red}\sim0.88$.
From the best fit, we obtain a disk accretion rate of $\sim$ 0.33\,$\dot{M}_{Edd}$ and a halo accretion rate of $\sim$ 0.16\,$\dot{M}_{\rm Edd}$ which are typical values for sources in HIMS (see e.g. Nath et al. 2024).
The shock is weak in strength ($R\sim1.07$) and stays at a distance of $\sim$ 122\,$r_s$ from the black hole.
From the best fit, the most probable mass of the black hole turns out to be $\sim$ 10.2\,$M_\odot$.
The cosine of the disk inclination angle is obtained to be $\sim$ 0.17, indicating a high disk inclination of $\sim 80^\circ$.
The best-fitted parameters are mentioned in Table 1, and the best-fitted spectrum is shown in Fig. 3a.

\begin{table*}%[!h]
	\centering
	\caption{Spectral Results of combined NICER (1-10\,keV) and {\it NuSTAR} (6-70\,keV) data.}
	\label{tab:table2}
	\begin{tabular}{lcr||lcr} % four columns, alignment for each
		\hline
  \hline
\multicolumn{2}{c}{\textsc{Tbabs(diskbb+thComp$\otimes$diskbb)}}  &   ~   & \multicolumn{2}{c}{\textsc{Tbabs({\tt reflect}$\otimes$TCAF)}}\\
   Tbabs &  $N_{\rm H}$ ($\times10^{22}~{\rm cm}^{-2}$) & 0.36$\pm$0.03 & Tbabs    & $N_{\rm H}$ ($\times10^{22}~{\rm cm}^{-2}$) & 0.50$\pm$0.03 \\ 
   diskbb &  $kT_{\rm in}$ (keV) & 0.31$\pm$0.01                         &  {\tt reflect} & $rel_{refl}$ &  11.4$\pm$4.1 \\
     &Norm$_{\rm DBB}$($\times10^{4}$)  & 33$\pm$13                     &          & $abund$ & 0.39$\pm$0.06 \\
   thComp$\otimes$diskbb &  $\Gamma_\tau$      & 1.76$\pm$0.004         &          & $Fe_{abund}$ & 0.034$\pm$0.009 \\
     &$kT_e$ (keV) & 5.63$\pm$0.04                                      &          & $cosIncl$ & 0.17$\pm$0.09 \\
     &$f_{\rm cov}$      & 0.99$\pm$0.05                                &  TCAF & $\dot{m}_d$ ($\dot{M}_{Edd}$)  & 0.33$\pm$0.01  \\
     &Norm$_{\rm ThComp}$ ($\times10^{4}$) & 43$\pm$18                  &       & $\dot{m}_h$ ($\dot{M}_{Edd}$)  & 0.16$\pm$0.01  \\
     &$\chi^2/DOF$ & 2332.37/815                                        &       & $M_{BH}$ ($M_\odot$) & 10.2$\pm$0.4 \\
   & ~ & ~    								& ~	& $X_s$ ($r_s$) & 122$\pm$2 \\
   & ~ & ~    								& ~ & $R$ & 1.07$\pm$0.01 \\
   & ~ & ~    								& ~ & Norm & 1478$\pm$24 \\
   & ~ & ~    								& ~ & $\chi^2/DOF$ & 717.23/811 \\
   \hline
\multicolumn{2}{c}{\textsc{Tbabs(diskbb+thComp$\otimes$diskbb+{\tt xillver})}}  & ~  &\multicolumn{2}{c}{\textsc{Tbabs(diskbb+{\tt relxill})}}\\
     Tbabs &   $N_{\rm H}$ ($\times10^{22}~{\rm cm}^{-2}$) & 0.25$\pm$0.02 &Tbabs&$N_{\rm H}$ ($\times10^{22}~{\rm cm}^{-2}$) &   0.30$\pm$0.03 \\   
     diskbb &$kT_{\rm in}$ (keV) & 0.39$\pm$0.01                            & diskbb&$kT_{\rm in}$  (keV)  &        0.39$\pm$0.03 \\  
     &Norm$_{\rm DBB}$($\times10^{4}$)  & 4$\pm$0.5                        &       &Norm($\times10^{4}$) &     13$\pm$6 \\  
     thComp$\otimes DiskBB$ &$\Gamma_\tau$      & 1.53$\pm$0.02            &{\tt relxill}&Index1               &     3.15$\pm$0.49 \\    
     &$kT_e$ (keV) & 20.06$\pm$6.31                                        &       &Index2               &     2.13$\pm$0.20 \\    
     &$f_{\rm cov}$      & 0.54$\pm$0.01                                   &       &R$_{\rm br}$             &      11.21$\pm$1.73 \\     
     &Norm$_{\rm thComp}$ ($\times10^{4}$) & 43$\pm$4                      &       &a                    &         0.96$\pm$0.15 \\    
   {\tt xillver}&$\Gamma$         &         1.53$\pm$0.08                        &       &Incl ($\theta$)      &          87$\pm$5.61 \\      
   &$A_{\rm Fe}$          &       1                                        &       &R$_{\rm in}$(ISCO)       & -1.53$\pm$0.77 \\            
   &$E_{\rm cut}$ (keV)   &       17.13$\pm$4.33                           &       &R$_{\rm out}$            &           1000.00$^a$ \\    
   &$log\xi$          &        3.62$\pm$0.09                               &       &$\Gamma$             &   1.44$\pm$0.04 \\      
   &Incl ($\theta$)   &        79.99$\pm$11                                &       &$log\xi$              &          0.69$\pm$0.03 \\      
   &refl$_{\rm frac}$       &          0.21$\pm$0.01                       &       &$A_{\rm Fe}$             &        1$^a$ \\        
    &Norm             &         0.25$\pm$0.09                              &       &$E_{\rm cut}$(keV)       &    17.87$\pm$0.76 \\    
    & $\chi^2/DOF$ & 703.32/810                                            &       &refl$_{\rm frac}$          &         1.72$\pm$0.03 \\      
   & ~ & ~									   &  ~  &Norm                 &       0.50$\pm$0.01 \\ 
   & ~ & ~									   &   ~ &      $\chi^2/DOF$ & 641.96/808 \\
   \hline
      \end{tabular}
     % \noindent{The errors are calculated using {\tt fit err} command and represent 90\% confidence level. $^a$ denotes fixed parameters.}
\end{table*}

%To make the model combination simple
To investigate the reflection component in detail, we again fit the combined spectra with Tbabs(diskbb+{\tt relxill}) model. We obtain similar values of $kT_{\rm in}$ and $N_{\rm H}$ values as in the previous results. We obtain the inner disk radius, R$_{\rm in}$ to be 1.53$\pm$0.77 ${\times}{\rm R}_{\rm ISCO}$ (ISCO: innermost (last) stable circular orbit). The inclination of the inner accretion disk gives high values ($\sim87\pm6$). We fixed the outer disk radius at 1000 r$_g$ (=$GM/c^2$) and iron abundance at solar value. The $\Gamma$ is now 1.44$\pm$0.04 and log$\xi\sim$ 0.69$\pm$0.03. The high energy cut-off is similar ($\sim17.87\pm0.76$\,keV) to the previous model fit. We obtain a high spin value ($\sim0.96$) for the best fit. The best-fit spectra retain a $\chi^2_{red}$ value of 0.79. The model parameters are presented in Table. 1 and the best-fit spectra are shown in Fig. 4b.

\section{Discussion and Concluding Remarks}

The first outburst of the Galactic BHXRB candidate Swift~J1727.8-1613 started on 24 August 2023 (MJD 60180).
After that, the source flux started to increase rapidly, reaching $\sim7.6$ Crab on 27 August 2023 (MJD 60183; Palmer \& Parsotan 2023).
In this paper, we report on the spectro-temporal analysis of data taken on 29 August 2023 (MJD 60185), when the source transitioned to the hard intermediate state from the initial hard state.
For the temporal analysis, we use data from NICER/XTI and construct the PDS in the $0.5-1$, $1-3$, $3-10$ and $0.5-10$\,keV energy bands and the cross-spectrum between the $0.5-3$\,keV and $3-10$\,keV light-curves.
From the cross-spectrum, we identified a soft lag of $-0.014 \pm 0.001$\,s seen at the QPO frequency ($\nu_{0}\pm\frac{\Delta\nu}{2}$). 
For spectral analysis, we use data from NICER/XTI and {\it NuSTAR}/FPM instruments to cover a broad $1-65$\,keV band and use various phenomenological and reflection models to fit the spectra.

One of the characteristic features of BHXRBs is the presence of low-frequency QPOs in the PDS, especially during the harder states of a BHXRB outburst.
It is well established that the variability in the Comptonized photon flux is responsible for producing the QPO peaks in the PDS, while the thermal disk emission shows very little variability (see e.g., Churazov et al. 2001).
Numerous models have been proposed to explain these oscillations, however, no models have yet been able to pinpoint the exact origin of these low frequency QPOs.
In the Two Component Advective Flow (TCAF; Chakrabarti \& Titarchuk 1995) model, LFQPOs are thought to be produced from the oscillation of the outer boundary of the CENtrifugal pressure dominated BOundary Layer (CENBOL) region.
In this model, accreting matter consists of two types: highly viscous and high angular momentum Keplerian flow and low viscous and low angular momentum sub-Keplerian flow.
The Keplerian flow accretes slowly due to high viscosity along the equatorial plane of the accretion disk, whereas the sub-Keplerian flow accretes rapidly due to its low viscosity above and below the disk plane.
This rapidly infalling sub-Keplerian flow undergoes a shock transition near the central black hole where the inward gravitational attraction on the matter becomes comparable with the outward centrifugal force.
Matter comes to a momentary halt at this shock boundary, and the post-shock matter becomes hot and the flow puffs up to form a torus like CENBOL.
The hot electrons inside this CENBOL inverse-Comptonizes the soft disk photons and produce hard Comptonized photons, essentially acting as the elusive `Compton cloud'.
The outer boundary of this CENBOL or the shock does not remain steady, rather it oscillates due to two main reasons: 

\begin{enumerate}
\item The Rankine-Hugoniot conditions which are necessary to sustain a steady shock (Chakrabarti 1989), are not satisfied (Ryu, Chakrabarti \& Molteni 1997),

%The satisfaction of the Rankine-Hugoniot conditions is necessary to sustain a steady shock (Chakrabarti 1989), is not satisfied (Ryu, Chakrabarti \& Molteni 1997), 

or,

\item The cooling timescale of the flow becomes comparable to the infall timescale, resulting in a resonance oscillation (Molteni, Sponholz \& Chakrabarti 1996).
\end{enumerate}

This oscillation of the shock modulates the Comptonized hard X-ray emission from the CENBOL, and the resulting variations of the hard X-ray intensity are observed as the quasi-periodic oscillations.
Oscillation of the CENBOL do not influence the soft X-ray emissions from the disk, which can explain the low variability in the disk emission.
From the top panel of Fig. 1, we can see that the $0.5-1$\,keV PDS has the minimum variability power, there is only one peaked noise component (QPO peak), and it becomes noisy after 6\,Hz, i.e., there is no significant source variability above this frequency other than the counting noise.
This shows that in the lower $0.5-1$\,keV energy band, the contribution of the Comptonized photons is less and the thermal photons dominate in this energy range.
If we look at the PDS of the next energy band, i.e. $1-3$\,keV, we can see that the total variability power has increased, there is significant source variability till a higher frequency of $\sim20$\,Hz, and a bump at $\sim1.8$\,Hz has appeared which looks like the harmonic of the $0.89$\,Hz QPO peak.
This indicates that the contribution from the Comptonized photons has increased in the $1-3$\,keV energy band.
Finally, if we look at the $3-10$\,keV energy band PDS, we can see that the variability power is maximum and the source flux shows significant variability for the entire frequency range of the PDS (up to $\sim50$\,Hz) and a clear harmonic peak has appeared at $\sim1.8$\,Hz.
This gives us a strong indication that the Comptonized photons are the responsible factor for the production of this QPO.

The less variable thermal disk photons enter the CENBOL region and undergo repeated inverse-Comptonization, which are then emitted as more variable high-energy photons that carry the signatures of the oscillation of the CENBOL.
However, due to the time spent during repeated scattering, the hard Comptonized photons generally reach the observer later than the soft disk photons. This is called hard lag.
This time lag between the photons in two energy bands can be estimated using the cross-spectrum between the light-curves in the two energy bands.
To estimate the time lag between the thermal disk photons and the Comptonized photons, we have constructed the cross-spectrum between the $0.5-3$\,keV light-curve where thermal radiation dominates and the $3-10$\,keV light-curve where Comptonized radiation dominates.
We determined the time lag at the QPO frequency to be $-0.014 \pm 0.001$\,s, which, contrary to the above picture, signifies a soft lag, i.e. soft photons lagging the hard photons.
Such soft lags have been observed previously in BHXRB sources having a high inclination (van den Eijnden et al. 2017) and prominent jet activity (Patra et al. 2019).
To infer the disk inclination of the system, we have performed spectral modeling of the reflection feature observed in the {\it NuSTAR} spectra.
From the spectral fit with both the {\tt xillver} and {\tt reflect} models, we have found the inclination of the system to be $\sim 80^\circ$, and from the fit with {\tt relxill} model, we have found the inclination of the system to be $\sim 87^\circ$. In such high inclination systems, effects such as disk reflection (Kara et al. 2013) and focusing of X-rays due to gravitational 
bending can impact the observed lag between hard and soft X-ray emission (Dutta \& Chakrabarti 2016).
A part of the hard X-rays from the CENBOL falls on the disk, is reprocessed and produces a relatively softer emission, which reaches the observer later than the hard X-rays.
The amount of such reflected soft emission increases as the inclination of the source increases, which can be a possible reason that soft lags are observed in high-inclination sources.
The high energy cut-off from our spectral fit with {\tt relxill} and the moderate distance of the shock from the black hole obtained from the spectral fit with TCAF suggests that the CENBOL is moderate in size. Hence, more high-energy photons emitted from the CENBOL get a chance to be reflected in the accretion disk before reaching the observer.
Moreover, hard X-rays emitted from the nearby region of the black hole suffer from gravitational bending and reach the observer faster than the soft emission from parts of the 
disk that are far away from the black hole, and this effect is also intensified along with the inclination of the source (see Fig. 7 of Dutta \& Chakrabarti 2016).
The moderate size of the CENBOL in this system also aids in this process, as emission from such a CENBOL naturally comes from regions close to the black hole, which gets affected by the gravitational bending effects. Thus, the combined effect of high inclination angle and size of the Comptonizing region, i.e. the CENBOL can contribute to produce the observed 
soft lags.

The values of $kT_{\rm in}$, $\Gamma$, as well as the disk and halo accretion rates obtained from our spectral study, indicate that the source is in the hard intermediate state, and 
strong jets and outflows are generally observed in this state. For Swift J1727.8-1613, a bright and continuous jet has been found to be present from polarimetry and radio 
observations (see Wood et al. 2024 and references therein). From studies of multiple high-inclination sources with strong jet emission, it has been observed that soft lags are 
higher when there is higher jet activity (Chatterjee et al. 2019, Patra et al. 2019). In the TCAF picture, it has been shown that the outflows originating from the CENBOL act as 
the base of the jet (Molteni et al. 1994). However, only a fraction of this outflow can achieve the required escape velocity and get emitted as the jet, and the other part of the 
outflow returns to the equatorial accretion flow (Kim et al. 2019).
This returning outflow generates a cooler and denser region, and this region intercepts the hard Comptonized X-rays and downscatters them to produce softer X-rays.
Soft X-rays produced in this way reach the observer later than the hard X-rays from the CENBOL, producing soft lags.
This effect is also intensified for high inclination sources, as gravitational bending increases the amount of hard X-ray emission in this case (Chatterjee et al. 2017b), which also increases the amount of downscattered soft X-rays from the returning flow region. 
The high inclination of the system is also supported by the observed P Cygni profile (Mu\~noz-Darias et al. 2016; S\'anchez-Sierras et al. 2023) for the candidate (Mata S\'anchez et al. 2024). In high-inclination systems, the line of sight is nearly parallel to the plane of the accretion disk. This orientation makes it easier to observe the blue-shifted absorption component caused by the material moving toward the observer, along with the emission from the receding material. The P Cygni profile tends to be more pronounced because the outflowing material crosses the observer's line of sight. Also, the presence of strong winds or jets in high inclination systems can lead to more noticeable P Cygni profiles, as the winds interact with the observer’s line of sight more directly. This behavior is similar to other highly inclined binaries such as V404 Cyg (Mata-Sanchez et al., 2018) and V4641 Sgr (Hare et al. 2020).
From the spectral fit with the TCAF model, the most probable mass of the compact object turns out to be $\sim$ 10.2\,$M_\odot$. However, to infer this value, we need to fit more samples of spectra, which will be done in the future and will be published elsewhere.

\section*{Acknowledgements}

We are thankful to the anonymous referee for his/her kind suggestions to improve the quality of the paper.
This work made use of NICER/XTI, and {\it NuSTAR}/FPM data supplied by the High Energy Astrophysics Science Archive Research Center (HEASARC) archive. S.K.N. acknowledges support from the SVMCM fellowship of the Government of West Bengal.
S.K.N. and D.D. acknowledge support from the ISRO-sponsored RESPOND project (ISRO/RES/2/418/17-18) fund. D.D. acknowledge the visiting research grant 
of National Tsing Hua University, Taiwan (NSTC 111-2811-M-007-066). D.C. acknowledge the grants NSPO-P-109221 of Taiwan Space Agency (TASA) and NSTC-112-2112-M-007-053 of National Science and Technology Committee of Taiwan. K.C. acknowledges support from the SWIFAR postdoctoral fellowship of Yunnan University. 
H.-K. C. is supported by NSTC of Taiwan under grant 112-2112-M-007-053.

\end{document}